\documentclass[sn-mathphys]{sn-jnl}
\jyear{2022}%
\theoremstyle{thmstyleone}%
\theoremstyle{thmstyletwo}%
\theoremstyle{thmstylethree}%
\usepackage{natbib}

\begin{document}

\title[Article Title]{Fusion cross section for the reaction $^{18}$O+$^{12}$C at 16.7 MeV/nucleon}

\author*[1,2]{\fnm{L.} \sur{Baldesi}}\email{lucia.baldesi@fi.infn.it}

\affil{for the NUCL-EX collaboration}
\affil{}

\affil*[1]{\orgdiv{Dipartimento di Fisica e Astronomia}, \orgname{Università di Firenze}, \postcode{50019} \state{Sesto Fiorentino}, \country{Italy}}

\affil[2]{\orgdiv{INFN - Sezione di Firenze}, \postcode{50019} \state{Sesto Fiorentino}, \country{Italy}}

\abstract{The absolute cross section for the complete fusion and the total reaction cross section as a function of the charge of the heavy residue have been indirectly estimated for the reaction $^{18}$O+$^{12}$C at 16.7 MeV/nucleon. Data have been collected with the GARFIELD+RCo setup at LNL (Italy). The mbarn/count normalization factor has been extracted using the statistical model HF$\ell$ (Hauser-Feshbach light).}


\maketitle

\section{Introduction}\label{sec1}

The study of light systems at low energy (7-12 MeV/nucleon) is one of the active research areas of the NUCL-EX collaboration. 
It is well known that at low beam energy the complete fusion is the main reaction channel for central collisions, almost exhausting the total reaction cross section. The complete fusion mechanism, characterized by the formation of a compound nucleus and its evaporation, is well described  by the statistical decay model based on the Hauser-Feshbach formalism \cite{bib7}. From the comparison between the data and a statistical model it was possible to put into evidence deviations from a statistical behaviour due to the presence of cluster structures \cite{bib8,bib9,bib10}. 
When the beam energy increases, the dynamic effects become relevant and other reaction mechanisms become significant, such as incomplete fusion and direct reactions.  In \cite{bib1} systematics of many available data on complete and incomplete fusion cross section as a function of $E_{CM}/A$, i.e. of the available energy the centre of mass (c.m.) divided by the total mass of the system, is presented. In particular, a parametrization describing the complete fusion fraction with respect to the total reaction cross section as a function of $E_{CM}/A$, independently of the reaction partners, is introduced. This parametrization shows that complete fusion becomes negligible at about $E_{CM}/A>7$ MeV, while it is still the dominant mechanism around 3-4 MeV. 

In this work the $^{18}$O+$^{12}$C light system at 16.7 MeV/nucleon, which corresponds to $E_{CM}/A=4$ MeV, will be discussed; events in which a heavy residue can be detected are selected, aiming to extract the fraction of complete fusion as a function of the charge of the residue. This goal will be achieved by comparing the experimental data with the results of a statistical simulation based on the Hauser-Feshbach approach and developed by the NUCL-EX collaboration to study light system \cite{bib3} (HF$\ell$ in the following).
For this reaction the systematics of \cite{bib1} predicts that the fusion cross section is around 17\% of the total reaction cross section, shared between complete (11\%) and incomplete (6\%) fusion while, according to \cite{bib5,bib11,bib12}, the reaction cross section can be estimated as (1.67 $\pm$ 0.07) barn.

The organization of the paper is the following: the experimental apparatus will be described in Section \ref{sec2}, while data analysis and results are discussed in Section \ref{sec3}.

\section{Experimental apparatus}\label{sec2}
\label{due}

In this experiment, a pulsed beam of $^{18}$O at 300 MeV delivered by the ALPI Linac of Laboratori Nazionali di Legnaro (LNL-Italy) with an average current of 0.1 pnA impinged on a $^{12}$C target with a thickness of 70 $\mu g\, cm^{-2}$ is used. 

Data were collected by means of the GARFIELD+RCo array \cite{bib2}.
This apparatus is characterized by a geometrical efficiency of almost 80$\%$ of the total solid angle and a good granularity (in fact it includes almost 300 $\Delta$E-E telescopes).
GARFIELD is a two-stage detector with azimuthal symmetry consisting of two MicroStrip Gas Chambers and 180 CsI(Tl) scintillators (arranged in 8 rings), readout by photodiodes, and it covers the angular range from 30$^{\circ}$ to 150$^{\circ}$ in polar angle. The RCo is placed at forward angles covering the polar range from 5$^{\circ}$ to 17$^{\circ}$. It is a three-stage detector including an Ionization Chamber (divided into 8 sectors) followed by 300$\mu m$-thick silicon strip detectors (8 strip for each sector) and as the last stage CsI(Tl) crystals (6 scintillators for each sector, read out by photodiodes). All the setup is equipped with digital electronics. GARFIELD is dedicated to the detection of Light Charged Particles (LCP) and Intermediate Mass Fragments (IMF), while the RCo is also aimed at the detection of the heavier fragments as the evaporation residues. A more complete description of the experimental apparatus can be found in \cite{bib2}.

\section{Experimental results and analysis}\label{sec3}

The main goal of this paper is to separate and estimate in the experimental data the contribution of the complete fusion channel, selecting, as first step, events which corresponds at least to a possible fusion (complete or incomplete) by looking to the presence of a evaporation residue candidate.

\begin{figure}[h]
\centering
\includegraphics[width=0.9\textwidth]{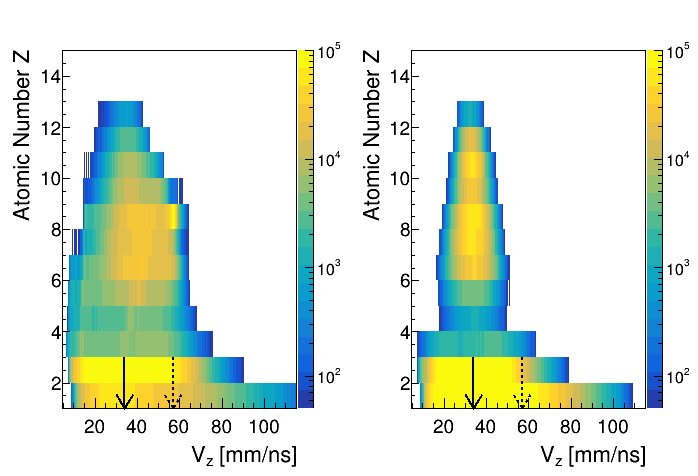}
\caption{Z vs. $v_z^{lab}$ correlations. Left panel: experimental data; right panel: HF$\ell$ simulation. Continuous arrow: center of mass velocity. Dotted arrow: projectile velocity.}
\label{fig1}
\end{figure}

To obtain this, it is useful to observe the correlation between the charge and the laboratory velocity component along the beam axis made without any events selection shown in Figure~\ref{fig1} (left panel). In this plot, a region beyond $Z=6$ and located close to the c.m. velocity ($v_{CM}=34.1$ mm/ns, continuous arrow) appears; the evaporation residues must be searched in this region. For $Z=8$ also an intensification at the beam velocity ($v_{p}=56.8$ mm/ns, dashed arrow) is evident, corresponding to the Quasi-Projectile QP of very peripheral reactions (for example, direct reactions). Light fragments are not very abundant, while the LCP's, with a wide velocity distribution extending up to 100 mm/ns (for $Z=1$), are more copious.

In order to estimate the complete fusion channel we used the statistical code HF$\ell$, described in \cite{bib3}, as it was optimized for the simulation of the decay of light compound nuclei keeping into account also the particle emission from discrete levels. As an input for the code, a source of $Z=14$, $A=30$, with excitation energy equal to 143.9 MeV and a triangular angular momentum distribution up to 31$\hbar$, was used. 
The simulated charge - lab velocity (component along the beam axis) correlation, obtained after filtering the model data via a software replica of the setup taking into account the geometrical coverage and the identification thresholds, is shown in Figure~\ref{fig1} right panel.
In this case, as expected, an intense spot centered around the $v_{CM}$ is obtained, accompanied by LCP and very few IMF.

\begin{figure}[h]
\centering
\includegraphics[width=1.0\textwidth]{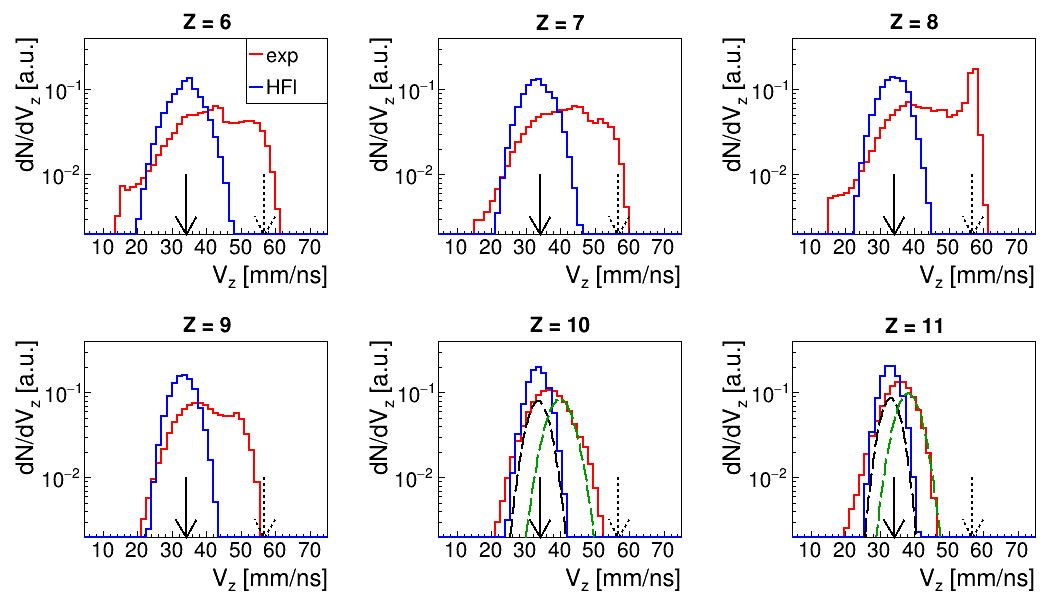}
\caption{Lab velocity distributions (beam axis component) for the heavy fragments. Red (blue) histograms: experimental (simulated) data. The distributions are normalized to their integral. Continuous arrow: $v_{CM}$; dotted arrow: $v_{p}$. For $Z=10$ and $Z=11$ as black and green dashed lines the two components of the gaussian fit (see text) are shown.}
\label{fig2}
\end{figure}

The results shown in the right panel of Figure~\ref{fig1} indicate that the complete fusion residue is mainly distributed among $Z=6$ and $Z=11$. So we will not consider $Z=12$ due to the low statistics. In Figure~\ref{fig2}, we show the experimental and simulated velocity distributions separately for each one of these Z values. The distributions are normalized to the integral to allow the spectra shape comparison. 

As already evident in Figure~\ref{fig1}, the experimental velocity spectra for $Z<10$ have a wider distribution than in the simulated case, thus suggesting that these fragments are not only evaporation residues coming from complete fusion reactions but they can be produced by other reaction mechanisms.

For $Z\geq10$ we note that the experimental spectra tend to be more similar to the simulated ones, but there is a contribution at a higher velocity that can originate from the incomplete fusion channel \cite{bib4}. 

In order to extract the cross section associated to the complete fusion channel for each Z, we need to find a normalization factor counts/mbarn. Due to the reaction grazing angle which is not covered by the apparatus, we don't have a direct measurement of the absolute cross section. Therefore we apply the alternative procedure here described.

In this self-consistent procedure the total cross section given by \cite{bib5,bib11,bib12} and the percentage of complete fusion given by the systematics (11\% for this reaction) have been used. 

The cross section for the complete fusion in this reaction is then equal to \(\sigma_{CF}=(184 \pm 42)\) mbarn, where the uncertainty comes from that on the percentage of complete fusion (2\% estimated graphically starting from Figure 10 of \cite{bib1}).

Assuming that HF$\ell$ predicts the correct evaporation residue distribution, we can use  the ratio of $\sigma_{CF}$ to $N^{HF\ell}_{tot}$, which is the number of fragments with $Z\geq6$, originally produced by the simulation without applying the filtering procedure, as a normalization factor for the model.

After evaluating the setup efficiency $\eta_i^{HF\ell}$  for each residue $i$ of charge $Z_i\geq6$, it is possible to estimate the expected cross section for the complete fusion in the experimental conditions:  

\begin{equation}
\sigma_{i}^{CF}=\sigma_{CF} \cdot BR_{i}^{HF\ell} \cdot \eta_{i}^{HF\ell}
\label{eq5}
\end{equation}
where BR$_{i}^{HF\ell}$ is the Branching Ratio for the channel $i$ (i.e. the ratio between the number of events in which a residue $Z_i$ is detected and the total number of filtered events in the simulation).

As already highlighted in Figure~\ref{fig2}, the complexity of the experimental spectra especially for $Z<10$ shows that different reaction mechanisms must be considered while we can assume that for $Z\geq10$ the significant processes are complete and incomplete fusion. Those two different contribution can be quantitatively estimated using a two gaussian fits, as already shown in \cite{bib4}. We decide to apply this technique for $Z=10$ and 11, where the statistics is higher and the contamination from direct reactions is negligible. The results of the two-gaussian fit on the velocity spectra is shown in Figure \ref{fig2} (dashed lines in panels for $Z=10,11$); the gaussian corresponding to the complete fusion component, shown as a black curve in  Figure \ref{fig2} (whose integral will be indicated as $G_i^{CF}$ for the residue of charge $Z_i$), was centered at $v_{CM}$, leaving the width as free parameter. 

Comparing the velocity spectra of experimental data and simulation, one should consider that the HF$\ell$ code is able to produce directly the velocity spectra in the C.M. system, while in the experimental data the velocity is not directly accessible because our apparatus can measure only the energy and the charge for particles as the evaporation residue candidate for $Z\geq10$.
To calculate experimentally the velocity it is necessary to know the mass of the particles. If this information is not known, the most probable mass value according to the Evaporation Attractor Line (EAL) \cite{bib6} is assigned to these particles. The mass distribution produced by the statistical code for Z$>$9 is different from the value suggested by the EAL, but to replicate the effects of the measuring apparatus on the model, the same mass as in the experimental case is attributed to these ions and the velocity is recalculated with it. It would be interesting to measure the masses and verify which of the two predictions is correct. In our analysis, it is only possible to choose one of the two masses and apply it consistently to both the model and the data. Fortunately, the systematic error on percentages of complete and incomplete fusion associated to the choice of the mass is $< 4\%$.

The experimental mbarn/count factors can then be obtained as \(\alpha_i=\frac{\sigma_{i}^{CF}}{G_i^{CF}}\), as shown in Table~\ref{tab1} for the residues with $Z=10$ and $Z=11$.

\begin{table}[h]
\begin{center}
\begin{minipage}{174pt}
\caption{Normalization factors for detected residues of $Z=10$ and $Z=11$}
\label{tab1}
\begin{tabular}{cc}
\toprule
 $\alpha_{10}$& $\alpha_{11}$ \\
 mbarn/count&mbarn/count\\
\midrule
(7.8 $\pm$ 1.8) $\cdot$ 10$^{-5}$  & (8.6 $\pm$ 2.0) $\cdot$ 10$^{-5}$
\\
\botrule
\end{tabular}
\end{minipage}
\end{center}
\end{table} 

If HF$\ell$ perfectly described all the characteristics of the complete fusion residue and the result of the two-Gaussian fits on the velocity distributions were reliable, the two normalization factors $\alpha_{i}$ would result equal. Indeed the values shown in Table~\ref{tab1} for $Z=10$ and $Z=11$ are compatible within errors. From the weighted average, we find the experimental normalization factor  $\alpha$ = (8.0 $\pm$ 1.9) $\cdot$ 10$^{-5}$ mbarn/count.

\begin{figure}[h]
\centering
\includegraphics[width=1.0\textwidth]{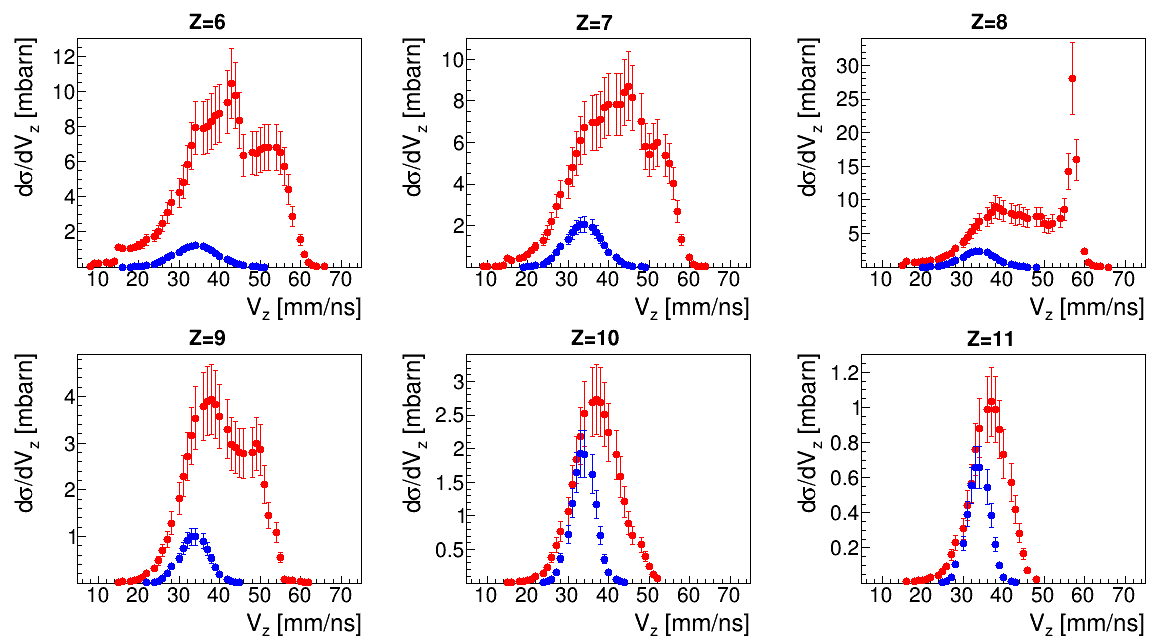}
\caption{Velocity distribution in mbarn for fragments with $5<Z<12$. The red points represent the total cross section and the blue ones the complete fusion cross section.}
\label{fig3}
\end{figure}

In Figure~\ref{fig3} the velocity spectra in absolute cross section, obtained applying the $\alpha$ normalization factor to those shown in Figure~\ref{fig2}, are presented, highlighting the complete fusion fraction (represented by the HF1 model in blue) with respect to the total experimental cross section (in red) for each Z.

\begin{figure}[h]
\centering
\includegraphics[width=1.0\textwidth]{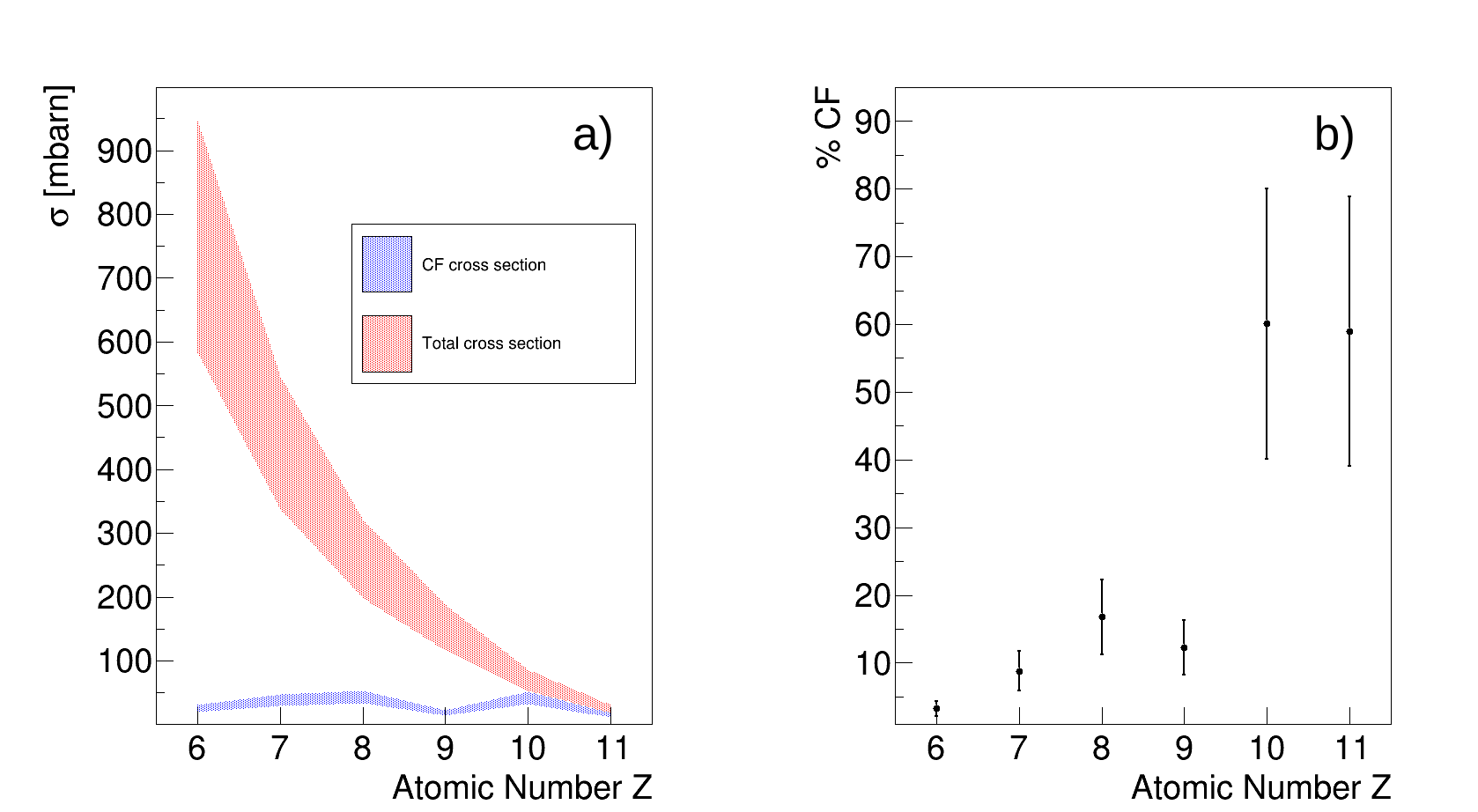}
\caption{On the left total (red) and complete fusion (blue) cross section vs Z of the residue.  On the right percentage of complete fusion vs Z. Data have been corrected for the efficiency of the setup.}
\label{fig4}
\end{figure}

We can then estimate the total (red) and complete (blue) fusion cross section, after the correction for the efficiency of the apparatus, for each residue with $5<Z<12$, as shown in Figure~\ref{fig4} (left panel). The experimental data are corrected for the efficiency of the setup using the code HIPSE \cite{bib13}, a semi-empirical model used by the nuclear physics community in order to simulate events which contains all the possible interaction mechanisms. In the case of complete fusion process, the efficiency correction has been performed using the HF$\ell$ code. 

Looking to Figure~\ref{fig4} panel a), one can note that the reaction cross section decreases significantly while the charge of the residue increases; while the contribution of the complete fusion show a more smooth trend, with a maximum around Z=8 and a even-odd effect between Z=9 and Z=10. 

In the right panel of Figure~\ref{fig4} the percentage of complete fusion as a function of the charge of the residue is shown. The percentage of complete fusion is almost negligible for the lightest residues, while it increases with $Z$ reaching values higher than 50\%. Comparing the area of the two gaussians corresponding to complete (black line) and incomplete (green line) for $Z=10$ and $Z=11$ in Figure~\ref{fig2} with the complete fusion percentage for such particles shown in Figure~\ref{fig4} panel b), one can note that there is an apparent inversion in the complete/incomplete percentage, while in Figure~\ref{fig2} incomplete fusion seems to be slightly dominant, in Figure~\ref{fig4} panel b) the complete fusion is dominant. This is due to the different efficiency factor used to correct the total and complete fusion cross section, as explained before. The relative errors associated with the percentage of complete fusion are almost constant and are about 20\%, mainly due to the error associated with the estimated complete fusion cross section.

\section{Summary and conclusions}\label{sec4}
In this work the total and complete fusion cross sections as a function of the charge of the residue for the reaction $^{18}$O+$^{12}$C at 16.7 MeV/nucleon have been estimated. The mbarn/count normalization factor has been extracted from measured charge and velocity distributions by means of a simulation based on the HF$\ell$ statistical model, particularly suited for light systems, thanks to the introduction of the particle decay from discrete levels and the use of the standard technique of the double gaussians fit to take into account the complete and incomplete fusion process. The model output has been also used to estimate the efficiency for the different residues by a yield comparison before and after a filtering procedure with a software replica of the setup.  The velocity spectra of the heavy residues ($Z=10,11$) have been fitted via a two-gaussian function, one of them fixed at v$_{CM}$ and representing the contribution of the complete fusion. From this experimental information it has been possible to derive the absolute contribution of the complete fusion for each Z-identified residue. It has been found that the complete fusion is quickly decreasing below Z=10 and it represents only around 10-20\% for residues with Z=7-9.


\begin{thebibliography}{99}
\addcontentsline{toc}{chapter}{Bibliografia}  

\bibitem{bib7} W. Hauser, H. Feshbach, Phys. Rev. 87, 366, (1952).

\bibitem{bib8} G. Baiocco et al. Phys. Rev. C 87, 054614 (2013).

\bibitem{bib9} L. Morelli et al. Phys. Rev. C 99, 054610 (2019).

\bibitem{bib10} M. Bruno et al. J. Phys. G: Nucl. Part. Phys. 46, 125101 (2019).

\bibitem{bib1} P. Eudes et al. Phys. Rev. C 90, 034609 (2014).

\bibitem{bib3} G. Baiocco: Towards a reconstruction of thermal properties of light nuclei from fusion-evaporation reaction. PhD Thesis, University of Bologna (2010).

\bibitem{bib5} S. k. Gupta et al. Zeit. Phys. A 317, 75 (1984).

\bibitem{bib11} S. Kox et al. Phys. Rev. C 35, 1678 (1987).

\bibitem{bib12} W.Q. Shen et al. Nuclear Physics A 491, 130–146 (1989).

\bibitem{bib2} M. Bruno et al. Eur. Phys. J. A 49, 128 (2013).

\bibitem{bib4} S. Pirrone et al. Phys. Rev. C 64, 024610 (2001).

\bibitem{bib6} R. J. Charity et al. Phys. Rev. C 58, 1073 (1998).

\bibitem{bib13} D. Lacroix et al. Phys. Rev. C 69, 054604 (2004).

\end{thebibliography}
\end{document}